\documentclass{JHEP3}
\usepackage{amsfonts}
\usepackage[centertags]{amsmath}
\usepackage{amssymb}
\usepackage{cite}
\usepackage{psfrag}
\usepackage{graphicx}
\usepackage{bbm,bm}
\usepackage{epsfig, multicol}
\allowdisplaybreaks
\title{The S-matrix of the Faddeev-Reshetikhin Model, Diagonalizability and $PT$ Symmetry}
\author{Ashok Das\thanks{Also at Saha Institute for Nuclear Physics,1/AF Bidhannagar, Calcutta 700064, India}\\
Department of Physics and Astronomy,\\
University of Rochester,Rochester, NY 14627-0171, USA\\
Email: \email{das@pas.rochester.edu}}

\author{A. Melikyan, V.O. Rivelles\\
Instituto de F\'{\i}sica\\
Universidade de S\~{a}o Paulo, 05315-970,\\
S\~{a}o Paulo, SP, Brazil\\
E-mail: \email{arsen@pas.rochester.edu},\ \email{rivelles@fma.if.usp.br}}
\date {}
\abstract{
We study the question of diagonalizability of the Hamiltonian for the
Faddeev-Reshetikhin (FR) model in the two particle sector. Although the two
particle S-matrix element for the FR model, which may be relevant for the
quantization of  strings on $AdS_{5}\times S^{5}$, has been calculated
recently using field theoretic methods, we find that the Hamiltonian for the
system in this sector is not diagonalizable. We trace the difficulty to the fact that the interaction
term  in the Hamiltonian violating Lorentz invariance leads to discontinuity conditions
(matching conditions) that cannot be satisfied. We determine the most
general quartic interaction Hamiltonian that can be diagonalized. This
includes the bosonic Thirring model as well as the bosonic chiral Gross-Neveu model
which we find share the same S-matrix. We explain this by showing, through a Fierz transformation, that these two models are in fact equivalent. In addition, we find a general quartic interaction Hamiltonian, violating Lorentz invariance, that can be diagonalized
with the same two particle S-matrix element as calculated by Klose and
Zarembo for the FR model. This family of generalized interaction Hamiltonians is not Hermitian, but is $PT$ symmetric. We show that the wave functions for this system are also $PT$ symmetric. Thus, the theory is in a $PT$ unbroken phase 
which guarantees the reality of the energy spectrum as well as the unitarity
of the S-matrix.}
\preprint{}
\begin{document}
\section{Introduction}

The classical integrability of the superstring on $AdS_{5}\times S^{5}$ \cite{Bena:2003wd,Kazakov:2004qf,Kazakov:2004nh,Beisert:2005bm,Beisert:2004ag,Arutyunov:2004yx,Alday:2005gi,Dorey:2006zj} has
led to a lot of interesting studies recently. Through $AdS/CFT$
correspondence, a lot is already known on the gauge theory side (for reviews see \cite{Minahan:2006sk,Swanson:2005wz,Plefka:2005bk,
Beisert:2004ry,Tseytlin:2004xa,Tseytlin:2003ii,Belitsky:2004cz,Zarembo:2004hp,Beisert:2007ds}  and references therein). However,
the string theory presents several technical difficulties, as a result of which, even
though it is believed that integrability should hold in the quantum theory,
quantizing the string remains an open question until now. It is worth
recalling here that the Green-Schwarz string can be described by a symmetric
space sigma model. The flat currents in this model \cite{Bena:2003wd,Das:2004hy,Das:2005hp} which define the basic
variables of the theory, like all sigma models, satisfy non-ultralocal
Poisson bracket structures (that involve derivatives of delta functions) \cite{Faddeev:1982rn,Faddeev:1987ph,Mikhailov:2005sy,Maillet:1985ek,Duncan:1989vg}.
This is one of the main difficulties in carrying out the quantization of
this model \cite{Mann:2005ab,Arutyunov:2005hd,Frolov:2006cc,Mikhailov:2005sy,Arutyunov:2004vx,Alday:2005jm,Hernandez:2006tk,Freyhult:2006vr}. Quantization of the model is, of course, absolutely crucial in 
understanding, say, the spectrum of the theory. And this still remains an open
question.

In the context of the principal chiral sigma model on $SU(2)$, Faddeev and
Reshetikhin \cite{Faddeev:1985qu} have suggested that the question of quantization may be carried
out in the following manner. They propose that the Hamiltonian for the
original sigma model as well as the non-ultralocal Poisson brackets may be replaced
by another Hamiltonian and a new Poisson bracket structure (that is
ultralocal) which lead to the same dynamical equations. One should
quantize this new system and then recover the original sigma model afterwards in
a certain limit. The idea is completely parallel to that contained in the method of 
Bethe ansatz in relativistic models (such as the massive Thirring model), where one studies the S-matrix elements by expanding the theory
around the wrong vacuum and then tries to go back to the true vacuum of the
theory where all the negative energy states are filled. The simpler (FR)
model proposed by Faddeev and Reshetikhin is quite important from this point
of view in connection with quantization of strings on $AdS_{5}\times S^{5}$ \cite{Kazakov:2004qf,Gromov:2006dh,Gromov:2006cq,Arutyunov:2004vx,Arutyunov:2006iu,Janik:2006dc,Beisert:2006zy,Schafer-Nameki:2004ik,Beisert:2003ys,Bellucci:2006bv,Freyhult:2006vr}.

Recently, Klose and Zarembo (KZ) \cite{Klose:2006dd} calculated the
S-matrix for a number of $1+1$ dimensional integrable models (see also \cite{Roiban:2006yc,Plefka:2006ze,Klose:2006zd,Klose:2007wq,Klose:2007rz}) using standard
field theoretic methods \cite{Thacker:1980ei,Thacker:1982rg,Thacker:1974kv,Thacker:1976vp} in a simple manner. The simplicity of their method arises from the fact that the calculations are carried out in the wrong
vacuum \cite{Korepin:1997bk,Gaudin:1983}. In this case, it is well known, for example, that in the two
particle sector, the contribution to the S-matrix comes only from the bubble
diagrams and if the system is integrable, all other scattering elements can
be related to the two particle S-matrix \cite{Zamolodchikov:1978xm,Staudacher:2004tk,Arutyunov:2006yd,Beisert:2006qh}. One of the models studied by KZ is
indeed the FR model, whose S-matrix element for the positive energy two
particle states has a simple form that reflects the violation of Lorentz
invariance present in the interaction Hamiltonian. This is interesting and,
in fact, is relevant as a first step in understanding the quantization of
the string itself. However, since the S-matrix element for the FR model is
calculated in the wrong vacuum, it is necessary, as a next step, to go to the true vacuum to
extract physical results \cite{Korepin:1979qq}. This can be carried out by diagonalizing the
(quartic) Hamiltonian of the theory in this sector.

In this paper, we study the question of diagonalization of the two particle
Hamiltonian for the FR model systematically. Surprisingly, we find that the
quartic Hamiltonian for the FR model cannot be diagonalized. The problem
arises because of the Lorentz violating term in the interaction Hamiltonian
which leads to boundary (matching) conditions that cannot be satisfied to
determine the wave function. This is rather puzzling given the nice S-matrix
result of KZ. However, our results are consistent with the results of KZ in
the following way. We find that, while the discontinuity cannot be matched
across the boundary to determine the wave function for the system, the extra
term violating the boundary condition is orthogonal to the positive energy
two particle state. Therefore, it drops out when the inner product with
positive energy states is taken and the discontinuity condition, in this case, yields the correct S-matrix obtained by
KZ. This is completely consistent with the field theoretic calculation of KZ, 
which involves only a calculation of matrix elements. This, therefore, leads to the
first important new feature that results from our analysis and which had not been
observed earlier in other integrable models. Namely, while diagonalization
of a system (Hamiltonian) leads to the S-matrix, having the S-matrix element (say, from a field theoretic calculation)
does not automatically imply diagonalizability of the system.

As a result of this lack of diagonalizability of the quartic Hamiltonian, we
then searched for and determined the most general quartic Hamiltonian in this
context for which the boundary conditions can be matched. The set of
potentials which can thus be diagonalized include the bosonic Thirring model \cite{Chowdhury:2004,Bhattacharyya:2004fg}
(which was known earlier to be integrable) as well as the bosonic chiral 
Gross-Neveu model (which to the best of our knowledge had not been studied
earlier). Both these models respect Lorentz invariance and, in fact, we find
that the two systems share the same S-matrix. Following this puzzling coincidence, we investigate and show, through a Fierz transformation, that the bosonic Thirring model and the bosonic chiral Gross-Neveu model are in fact equivalent. In addition, we determine 
a general quartic Hamiltonian, violating  
Lorentz invariance, which can be diagonalized and leads to the same S-matrix as
calculated by KZ for the FR model. We emphasize here that a field theoretic calculation with this generalized potential (interaction vertex) would yield the same S-matrix element as for the FR model. This is indeed quite interesting and
another important result of our analysis, namely, different potentials (interaction vertices) can lead to the same S-matrix element in a field theoretic calculation. Furthermore, we find that in spite
of the fact that the spectrum of this generalized family of Hamiltonians
is real and  the S-matrix is unitary, the Hamiltonian is not Hermitian.
On closer analysis, we find that the new family of Hamiltonians is, in fact, 
$PT$ symmetric \cite{Bender:1998gh,Dorey:2001hi,Bender:2002vv} (for reviews see \cite{Bender:2007nj,Bender:2005tb} and references therein). We show that the theory is in the $PT$ unbroken phase which guarantees the reality of the spectrum as well as the
unitarity of the S-matrix. This is the third important result of our
analysis which identifies the relevance of $PT$ symmetry with this integrable system. This, of course, still leaves us with the interesting question of why
the FR model cannot be diagonalized and this is presently under study.

Our paper is organized as follows. In section {\bf 2}, we give a brief review of
the Faddeev and Reshetikhin model and its relevance to the 
regularization of the ambiguities in the current algebra of the $SU(2)$ principal
chiral model. In section {\bf 3},  we recapitulate the  relation of the FR model to the string sigma
model, in particular, within  the context of the  $R\times S^{3}$ subsector of $AdS_{5}\times S^{5}$ background for simplicity. Here we also describe briefly the field theoretic calculation by KZ of the 
S-matrix for this model. In section {\bf 4},  we demonstrate the non-diagonalizability
of the corresponding Hamiltonian in the operator formalism and try to make connection with the field theoretic calculation. In section {\bf 5}, we analyze the underlying quantum mechanical system to understand the difficulty in more detail. We show that the term in the interaction Hamiltonian, violating Lorentz invariance, leads to matching conditions that cannot be satisfied. Here we also make connection with the field theoretic results. In section {\bf 6}, we present  the general quartic Hamiltonian that can be diagonalized and determine the S-matrix associated with this system. Various special cases are studied here. In particular, we show that the bosonic Thirring interaction is equivalent to the bosonic chiral Gross-Neveu interaction. In addition, we determine a general Hamiltonian, violating Lorentz invariance, that can be diagonalized and has the same S-matrix as that calculated by KZ for the FR model. We show that this family of generalized Hamiltonians is not Hermitian, but is $PT$ symmetric. By studying the transformation properties of the wave functions of the system, we show that the system is in the $PT$ unbroken phase which guarantees the reality of the spectrum as well as the unitarity of the S-matrix. In section {\bf 7}, we give a brief summary of our results.

\section{The Faddeev-Reshetikhin model}

The Lagrangian for the $SU(N)$ principal chiral model is written in the
form: 
\begin{equation*}
L=\frac{1}{2\gamma }\int dx\ \eta ^{\mu \nu }J_{\mu }^{a}J_{\nu }^{a},\quad
\mu ,\nu =0,1,
\end{equation*}
where $J_{\mu }=-g^{-1}\partial _{\mu }g=J_{\mu }^{a}t^{a}$, $
[t^{a},t^{b}]=f^{abc}t^{c};\ a,b,c=1,2,\cdots N^{2}-1,$ 
$g\in SU(N).$ Here $\gamma $ is a constant and throughout the paper we use
the Bjorken-Drell metric which, in $1+1$ dimensions, has the diagonal form $
\eta ^{\mu \nu }=(+,-)$. From the definition, we see that the current $
J_{\mu }^{a}$ is a pure gauge and, therefore, satisfies a zero curvature
condition in addition to the dynamical equation. Thus, it can be shown that the (current) variables of the theory  satisfy the equations: 
\begin{eqnarray}
&&\partial _{\mu }J^{\mu ,a}=0,  \notag \\
&&\partial _{\mu }J_{\nu }^{a}-\partial _{\nu }J_{\mu }^{a}+f^{abc}J_{\mu
}^{b}J_{\nu }^{c}=0.  \label{eqns}
\end{eqnarray}
We can choose $J_{1}^{a}$ to be the dynamical variable and carry out the
Hamiltonian analysis for the system which leads to the following Poisson
bracket structures for the dynamical variables of the theory: 
\begin{eqnarray}
\{J_{0}^{a}(x),J_{0}^{b}(y)\} &=&\gamma f^{abc}J_{0}^{c}(x)\delta (x-y) 
\notag \\
&&  \notag \\
\{J_{0}^{a}(x),J_{1}^{b}(y)\} &=&\gamma f^{abc}J_{1}^{c}(x)\delta
(x-y)-\gamma \delta ^{ab}\partial _{x}\delta (x-y)  \label{pcmalgebra} \\
&&  \notag \\
\{J_{1}^{a}(x),J_{1}^{b}(y)\} &=&0  \notag
\end{eqnarray}
Because of the presence of terms involving derivatives of delta functions,
the Poisson bracket algebra becomes non-ultralocal. (We remark here parenthetically that the Poisson bracket for the nonlinear sigma model is not unique \cite{Brunelli:2002fh}. However, in any form, there always exist some derivatives of delta function, making them non-ultralocal.) This leads to
ambiguities in the calculation of the basic algebra of transition matrices
and renders inapplicable the standard procedure of quantization for such
systems.

The Faddeev-Reshetikhin (FR) model studies the quantization of the $SU(2)$
principal chiral model for which $a,b,c=1,2,3,\ f^{abc}=-\varepsilon ^{abc}$
and the generators are related to the Pauli matrices, namely, $t^{a}=i\sigma ^{a}/2$. In this case, the
basic variables of the theory can also be written as (3 dimensional)
vectors in the internal space. Thus, defining 
\begin{eqnarray}
\vec{S}_{+} &=&\frac{1}{4\gamma }(\vec{J}_{0}+ \vec{J}_{1}),  \notag \\
&&  \label{Splusminus} \\
\vec{S}_{-} &=&\frac{1}{4\gamma }(\vec{J}_{0}-\vec{J}_{1}),  \notag
\end{eqnarray}%
we note that in these variables,  the equations in \eqref{eqns} take the forms 
\begin{equation}
\partial _{t}\vec{S}_{\pm }\mp \partial _{x}\vec{S}_{\pm }\pm 2\gamma \vec{S}
_{+}\times \vec{S}_{-}=0,  \label{eqn}
\end{equation}
where we have identified  $\vec{S
}_{\pm }=\left( S_{\pm}^{1},S_{\pm}^{2},S_{\pm}^{3}\right) $.

It is the term involving the derivative of delta function in 
\eqref{pcmalgebra} that leads to difficulty in quantization of the system.
The proposal of Faddeev and Reshetikhin is, therefore, to introduce the new Poisson bracket  
\begin{eqnarray}
\{S_{\pm }^{a}(x),S_{\pm }^{b}(y)\} &=&-\varepsilon ^{abc}S_{\pm
}^{c}(x)\delta (x-y),  \notag \\
&&  \label{newpoisson} \\
\{S_{+}^{a}(x),S_{-}^{b}(y)\} &=&0,  \notag
\end{eqnarray}
which does not contain any non-ultralocal terms and a new Hamiltonian which would give rise to the same equations as in \eqref{eqn}. It follows now from 
\eqref{newpoisson} that the consistency of these relations requires
\begin{equation}
| \vec{S}_{+}| =| \vec{S}_{-}| =\lvert
S\rvert .  \label{Virasoro1}
\end{equation}
On the other hand, for the system to
have the same equation of motion \eqref{eqn}, the Hamiltonian of the system
must modify as well (along with the Poisson bracket) and the new Hamiltonian
is given by 
\begin{equation}
\mathcal{H=-(P}_{S_{+}}-\mathcal{P}_{S_{-}}\mathcal{)+}2\gamma \int dx\ \vec{
S}_{+}\cdot \vec{S}_{-},  \label{FRHam}
\end{equation}
where $\mathcal{P}_{S_{+}}$ and $\mathcal{P}_{S_{-}}$ are the momenta associated with the two variables, with the explicit forms:
\begin{eqnarray}
\mathcal{P}_{S_{+}} &=&\int \frac{S_{+}^{1}\partial
_{x}S_{+}^{2}-S_{+}^{2}\partial _{x}S_{+}^{1}}{| S|
+S_{+}^{3}},  \notag \\
&&  \label{momenta2} \\
\mathcal{P}_{S_{-}} &=&\int \frac{S_{-}^{1}\partial
_{x}S_{-}^{2}-S_{-}^{2}\partial _{x}S_{-}^{1}}{| S|
+S_{-}^{3}}.  \notag
\end{eqnarray}
It is easy to check using \eqref{newpoisson} that 
\begin{equation}
\{\mathcal{P}_{\vec{S}_{+}},\vec{S}_{+}\}=-\partial _{x}\vec{S}_{+},\quad \{
\mathcal{P}_{\vec{S}_{-}},\vec{S}_{-}\}=-\partial _{x}\vec{S}_{-},
\end{equation}
so that the dynamical equations \eqref{eqn} arise as Hamiltonian equations
with the modified Hamiltonian \eqref{FRHam} as well as the modified Poisson brackets 
\eqref{newpoisson}.

The dynamical system described by \eqref{eqn} has an infinite set of
conserved charges associated with it and the Hamiltonian (\ref{FRHam}) can
be determined from the low order charges. To see how the infinite set of
conserved charges arise, let us note that the dynamical equations \eqref{eqn}
can be obtained from the zero curvature (flat) condition associated with a
one parameter family of currents 
\begin{equation}
\partial _{t}\vec{S}_{+}(\lambda )-\partial _{x}\vec{S}_{-}(\lambda )+2\vec{
S}_{+}(\lambda )\times \vec{S}_{-}(\lambda )=0,  \label{UV}
\end{equation}
where $\lambda$ is a constant (spectral) parameter and we have defined
\begin{equation}
\vec{S}_{\pm }(\lambda )=\frac{\vec{S}_{+}}{\lambda -a}\pm \frac{\vec{S}_{-}
}{\lambda +a},  \label{Jplusminus}
\end{equation}
with $a\equiv 1/\gamma .$ The transfer matrix
\begin{equation*}
T(x,y,\lambda )=\mathcal{P}\exp \left( \overset{x}{\underset{y}{\int }}dz\
S_{+}(z,\lambda )\right) ,
\end{equation*}
where $\mathcal{P}$ denotes path ordering, can be decomposed in the standard
form
\begin{equation}
T(x,y,\lambda )=\left[ I+W(x,y,\lambda )\right] \exp \left( Z(x,y,\lambda
)\right) \left[ I+W(x,y,\lambda )\right] ^{-1},  \label{Ttransdecomp}
\end{equation}
where $Z(x,y,\lambda )$and $W(x,y,\lambda )$ are respectively diagonal and
anti-diagonal matrices. The monodromy matrix, which contains all the
conserved charges of the theory, is defined as (we assume that the theory is defined for $-L\leq x\leq L$ with the continuum limit obtained for $L\rightarrow \infty$)
\begin{equation}
T_{L}\left( \lambda \right) = T(L,-L,\lambda )  \label{monmat2}
\end{equation}
One can now obtain the local charges by expanding the monodromy matrix
around the two poles in (\ref{Jplusminus}) which leads to two sets of
conserved charges $\varphi _{L}^{\pm (n)}$ from the series:
\begin{equation}
\varphi _{L}\left( \lambda \right) =\underset{n}{\sum }\left( \lambda \pm
a\right) ^{n}\varphi _{L}^{\pm (n)},  \label{phidecomp}
\end{equation}
where the generating functional $\varphi _{L}\left( \lambda \right) $ is
defined through the relation
\begin{equation}
{\rm Tr}\left[ T_{L}\left( \lambda \right) \right] =2\cos \varphi _{L}\left(
\lambda \right).  \label{defphi}
\end{equation}
For the FR model, the decomposition (\ref{phidecomp}) has the 
explicit form:
\begin{equation}
\varphi _{L}^{\pm }\left( \lambda \right) =\frac{1}{\lambda \pm a}\varphi
_{L}^{\pm (-1)}+\varphi _{L}^{\pm (0)}+\left( \lambda \pm a\right) \varphi
_{L}^{\pm (1)}+...  \label{phiexplicit}
\end{equation}
In this series the trivial conserved charges correspond to the conservation of total spin: 
\begin{equation}
\varphi _{L}^{\pm (-1)}=\overset{L}{\underset{-L}{\int }}|
S| dx  \label{zerocharge}
\end{equation}
The first non-trivial charges have the form: 
\begin{equation*}
\varphi _{L}^{\pm (0)}=-\mathcal{P}_{S_{\pm }}\pm \frac{1}{a}\overset{L}{
\underset{-L}{\int }}\frac{\vec{S}_{+}\cdot \vec{S}_{-}}{|
S| }dx
\end{equation*}
where $\mathcal{P}_{S_{\pm }}$ are the momenta defined in (\ref{momenta2}).

To summarize, therefore, the FR proposal is to modify the Poisson bracket (without any non-ultralocal term)  as well as the Hamiltonian of the theory such that the same dynamical equations result. The theory can now be quantized. The idea here is that the term with the derivative of the delta function in the original Poisson brackets (say, in \eqref{pcmalgebra}) is an anomalous term (can be thought of as an anomaly) which may arise in some limiting manner after quantization. 

\section{String Sigma Model.}

As we have mentioned earlier, in the Green-Schwarz formulation, the
superstring on $AdS_{5}\times S^{5}$ can be described as a symmetric space
sigma model \cite{Metsaev:1998it}. Let $g$ denote an element of the graded group $PSU(2,2|4)$
(namely, it represents a map from the worldsheet onto the graded group $
PSU(2,2|4)$). Then, the left-invariant current associated with the group $
J=-g^{-1}\mathrm{d}g$ can be decomposed in terms of the superLie algebra
elements of $psu(2,2|4)$, which has a natural $\mathbf{Z}_{4}$ symmetry
(automorphism), as 
\begin{equation}
J=-g^{-1}\mathrm{d}g=H+P+Q^{1}+Q^{2}.  \label{curdecomp1}
\end{equation}
Here $H$ denotes elements of the maximal (non-compact) bosonic subalgebra $
so(4,1)\times so(5)$, while $P$ denotes elements of the bosonic complement
and $Q^{1},Q^{2}$ represent the Grassmann elements of the superalgebra under
the $\mathbf{Z}_{4}$ grading. In terms of these variables, the string sigma
model action on $AdS_{5}\times S^{5}$ can be written as 
\begin{equation}
S=\frac{1}{2}\int \mathrm{str}\left( P\wedge {}^{\ast }P-Q^{1}\wedge {Q^{2}}
\right) ,  \label{newaction}
\end{equation}
where \textquotedblleft str" stands for supertrace, ``$*$" denotes the Hodge dual operation and the second term in 
\eqref{newaction} represents a fermionic Wess-Zumino term. Introducing the
one parameter family of flat currents \cite{Das:2004hy} $\hat{J}(\lambda )\equiv -\hat{g}
^{-1}(\lambda )\mathrm{d}\hat{g}(\lambda )$, where $\lambda $ is a constant
spectral parameter,
\begin{equation}
\hat{J}(\lambda )=H+\frac{1+\lambda ^{2}}{1-\lambda ^{2}}\ P+\frac{2\lambda 
}{1-\lambda ^{2}}\ {}^{\ast }P+\sqrt{\frac{1}{1-\lambda ^{2}}}\ Q+\sqrt{
\frac{\lambda ^{2}}{1-\lambda ^{2}}}\ Q^{\prime },  \label{flatcurrent}
\end{equation}
such that $\hat{J}(\lambda =0)=J$, it is easy to check that the vanishing
curvature condition for this current 
\begin{equation}
\mathrm{d}\hat{J}-\hat{J}\wedge \hat{J}=0,  \label{zerocurv2}
\end{equation}
leads to the equations of motion for the system following from the action 
(\ref{newaction}). Here we have defined $Q = Q^{1}+Q^{2}, Q^{\prime} = Q^{1}-Q^{2}$.

In \cite{Das:2004hy,Das:2005hp} we have calculated explicitly the algebra $\{\hat{J}
_{1}(\sigma ,t)\overset{\otimes }{,}\hat{J}_{1}(\sigma ^{\prime
},t)\}$, and have shown that this Poisson bracket algebra contains the
problematic $\partial _{\sigma }\delta (\sigma -\sigma ^{\prime })$ term in
general. Therefore, the FR approach may be relevant in understanding the
quantization of this system. The presence of the Wess-Zumino term in the
action in (\ref{newaction}) can also be incorporated into the proposal of FR
\cite{Faddeev:1985qu}, by weakening the constraint \eqref{Virasoro1} so that the lengths $| \vec{S}_{+}|$  and $|
\vec{S}_{-}| $ are not equal. Thus, the Faddeev-Reshetikhin proposal is a promising
and interesting scheme that may potentially allow one to proceed with the
quantization of the string sigma model on $AdS_{5}\times S^{5}$ and deserves
to be investigated. However, to keep the discussion parallel to the $SU(2)$
principal chiral model of the last section, we will restrict ourselves here
to the string sigma model on the $R\times S^{3}$ subsector of $AdS_{5}\times
S^{5}$. Unlike the principal chiral model in flat space, here the equations
of motion need to be supplemented by the Virasoro constraints. Essentially,
what this means is that, unlike in the case of the principal chiral model,
in the case of the string, the constraint (\ref{Virasoro1}) does not have to
be set by hand.

In fact, an important progress has been made in this direction recently by Klose and Zarembo
\cite{Klose:2006dd}, where the S-matrix calculation has been carried out 
for the FR model (among several other models). The calculation involves
summing up a particular set of Feynman diagrams, the bubble diagrams, for
the two-particle scattering, which are the relevant diagrams for
calculations performed in the wrong vacuum. Using the factorization property
of the S-matrix of an integrable system \cite{Zamolodchikov:1978xm}, one can then write
down the S-matrix for general $N$-particle scattering. The corresponding
Bethe Ansatz for the theory can then be obtained by imposing relevant
boundary conditions. Since this calculation is rather relevant from our
point of view, here we briefly review their main results directing the reader to \cite{Klose:2006dd} for technical details.

The action for strings propagating on $R\times S^{3}$ subsector of $
AdS_{5}\times S^{5}\ $has the form:
\begin{equation}
S=-\frac{\sqrt{\lambda ^{\prime }}}{4\pi }\int d\tau d\sigma\ \eta^{\mu
\nu }\left[ \frac{1}{2}Tr(J_{\mu }J_{\nu })+\partial _{\mu }X^{0}\partial
_{\nu }X^{0}\right] ,  \label{rs3}
\end{equation}
where $\sqrt{\lambda^{\prime}/2\pi}$ represents the string tension,  the current $J=-g^{-1}dg\in su(2)$, and $g$ is an element
of the standard map $S^{3}\longrightarrow SU(2)$
\begin{equation}
g=\left( 
\begin{array}{cc}
X^{1}+iX^{2} & X^{3}+iX^{4} \\ 
-X^{3}+iX^{4} & X^{1}-iX^{2}
\end{array}
\right).  \label{gmap}
\end{equation}
The action \eqref{rs3} is written in the conformal gauge
\begin{equation}
\sqrt{-g} g^{\mu\nu} = \eta^{\mu\nu},\label{confgauge}
\end{equation}
and introducing the light-cone coordinates $\sigma _{\pm }=\frac{1}{2}(\tau \pm
\sigma )$, $\partial _{\pm }=\partial _{\tau }\pm \partial _{\sigma },\ $
we note that the Virasoro constraints of the theory take the form:
\begin{equation}
Tr(J_{\pm }^{2})=-2\left( \partial _{\pm }X^{0}\right) ^{2}.
\label{virasorolightcone}
\end{equation}
Furthermore, if we use the parameterization
\begin{equation}
X^{0} =\kappa \tau ,  
\end{equation}
where $\kappa$ denotes a constant, the Virasoro constraints in \eqref{virasorolightcone} take the simpler form:
\begin{equation}
Tr(J_{\pm }^{2})=-2\kappa ^{2}.  \label{virasoro3}
\end{equation}

To make connection with the $SU(2)$ principal chiral model considered by
Faddeev and Reshetikhin, let us identify: 
\begin{eqnarray}
J_{\pm } &=&i\kappa \vec{S}_{\pm }\cdot \vec{\sigma },
\notag \\
&&  \label{spinparam} \\
\vec{S}_{\pm } &=&(S_{\pm }^{1},S_{\pm }^{2},S_{\pm }^{3}), 
\notag
\end{eqnarray}
so that the Virasoro constraints and the equations of motion of the theory, in terms of the
spin variables $\vec{S}_{\pm }$ become
\begin{eqnarray}
\vec{S}_{\pm }^{2} &=&1  \notag \\
&&  \notag \\
\partial _{\mp }\vec{S}_{\pm }\mp \kappa \vec{S}_{+}\times \vec{S}_{-} &=&0,
\label{spinequations}
\end{eqnarray}
which can be compared with \eqref{eqn} and \eqref{Virasoro1}. 
Thus, the analogy with the FR model is now clear and one can follow the FR proposal to quantize the system. The action corresponding
to the Hamiltonian (\ref{FRHam}) takes the form:
\begin{equation}
S=\int d^{2}x\left[ -\left( C_{+}(\vec{S}_{-})+C_{-}(
\vec{S}_{+})\right) -\frac{\kappa }{2}\vec{S}_{+}\cdot 
\vec{S}_{-}\right] ,  \label{KZ1}
\end{equation}
where $C_{+}(\vec{S}_{-})$ and $C_{-}(\vec{S}_{+})$
are the Wess-Zumino terms which have the explicit forms
\begin{eqnarray}
C_{+}(\vec{S}_{-}) &=&-\frac{1}{2}\overset{1}{\underset{0}{\int }}
d\xi \ \varepsilon ^{abc}S_{-}^{a}\partial _{\xi }S_{-}^{b}\partial
_{+}S_{-}^{c},  \notag \\
&&  \label{WZ1} \\
C_{-}(\vec{S}_{+}) &=&-\frac{1}{2}\overset{1}{\underset{0}{\int }}
d\xi \ \varepsilon ^{abc}S_{+}^{a}\partial _{\xi }S_{+}^{b}\partial
_{-}S_{+}^{c},  \notag
\end{eqnarray}
with the boundary conditions
\begin{equation}
\vec{S}_{\pm }(\tau ,\sigma ,\xi =1)=\left( \vec{S}
_{\pm }\right) _{0}=\mathrm{const},  \label{WZboundary1}
\end{equation}
\begin{equation}
\vec{S}_{\pm }(\tau ,\sigma ,\xi =0)=\vec{S}_{\pm
}(\tau ,\sigma ).  \label{WZboundary2}
\end{equation}

Rewriting the Wess-Zumino terms (\ref{WZ1}) in the local, non-covariant
forms leads to the expression in (\ref{momenta2}) for the momenta of the theory, 
if one chooses $| S| =1$. Solving the constraint $
\vec{S}_{\pm }^{2}=1$ and introducing 
\begin{equation}
\phi _{\pm }=\frac{S_{\pm }^{1}+iS_{\pm }^{2}}{\sqrt{2}\sqrt{(1+S_{\pm }^{3})
}},  \label{phi}
\end{equation}
the action \eqref{KZ1} can be written in the unconstrained form:
\begin{eqnarray}
S &=&\int d^{2}x\left[ \frac{i}{2}\left( \phi _{-}^{\ast }\partial _{+}\phi
_{-}-\phi _{-}\partial _{+}\phi _{-}^{\ast }\right) +\frac{i}{2}\left( \phi
_{+}^{\ast }\partial _{-}\phi _{+}-\phi _{+}\partial _{-}\phi _{+}^{\ast
}\right) +\kappa \left( | \phi _{+}| ^{2}+| \phi
_{-}| ^{2}\right) \right.  \notag \\
&&  \label{KZ11} \\
&&\left. -\kappa \sqrt{(1-| \phi _{+}| ^{2})(1-|
\phi _{-}| ^{2})}(\phi _{+}^{\ast }\phi _{-}+\phi _{-}^{\ast }\phi
_{+})-2\kappa | \phi _{+}| ^{2}| \phi
_{-}| ^{2}\right].  \notag
\end{eqnarray}
Finally, introducing the two component bosonic spinor
\begin{equation*}
\phi =\binom{\phi _{-}}{\phi _{+}} = \binom{\phi _{1}}{\phi _{2}},
\end{equation*}
the action (\ref{KZ11}) can be recast into the following compact form for
terms up to the order $\phi ^{4}$ (quartic order):
\begin{equation}
S=\int d^{2}x\left[ i\overline{\phi }\gamma ^{\mu }D_{\mu }\phi -m\overline{
\phi }\phi -g\left( \overline{\phi }\gamma ^{\mu }\phi \right) \left( 
\overline{\phi }\gamma _{\mu }\phi \right) +O(\phi ^{6})\right] ,
\label{KZmain}
\end{equation}
where $D_{0}=\partial _{0}-im-ig\overline{\phi }\phi $; $D_{1}=\partial _{1}=\partial_{x}$
; $m=\kappa $; $g=\frac{\kappa }{2}.$ The non-covariance of the model is
hidden in the field dependent chemical potential in the definition of the
covariant derivative, which when written out explicitly would lead to a term
in the quartic interaction Hamiltonian that violates Lorentz invariance. We note here that throughout the paper, the interactions are assumed to be normal ordered, although we do not explicitly write the normal ordering symbol for simplicity.

We do not go into the details of Klose and Zarembo's calculations which are
explained nicely in \cite{Klose:2006dd}. Rather, we would like to summarize the essential
features of their calculation which would be relevant for our discussion.
First of all, the theory in \eqref{KZmain} has both positive and negative
energy solutions, as any free relativistic theory would have. For free particles, they satisfy the momentum space equations
\begin{equation}
(k\!\!\!\slash - m) u (k) = 0 = (k\!\!\!\slash + m) v (k),\label{covariant}
\end{equation}
and have the explicit forms
\begin{equation}
u (k) = \sqrt{m} \left(\begin{array}{c}
e^{\frac{\beta}{2}}\\
e^{-\frac{\beta}{2}}
\end{array}\right),\quad v (k) = \sqrt{m}\left(\begin{array}{r}
-e^{\frac{\beta}{2}}\\
e^{-\frac{\beta}{2}}
\end{array}\right).\label{covariant1}
\end{equation}
Here we have used the rapidity variable defined as
\begin{equation}
k = m\sinh \beta,\quad E_{k} = \sqrt{k^{2}+m^{2}} = m \cosh \beta,\label{rapidity}
\end{equation}
to parameterize the solutions and we have chosen the Lorentz invariant normalization 
\begin{equation}
\bar{u} (k) u (k) = 2m = - \bar{v} (k) v (k),\label{normalization}
\end{equation}
with $\bar{u} (k) = u^{\dagger} (k) \gamma^{0}$.

If we quantize the theory in the wrong vacuum:
\begin{equation}
\phi (x)| 0\rangle =0,  \label{vacuum}
\end{equation}
then the field can be decomposed completely in terms of annihilation operators as
\begin{equation}
\phi (x) = \int \frac{dk}{\sqrt{4\pi E_{k}}}\left(e^{-ik\cdot x} u (k) a (k) + e^{ik\cdot x} v (k) b (k)\right),
\end{equation}
Here we have identified $k^{0}=E_{k}, k^{1}=k$ and using the (nontrivial) equal-time commutation relation for the  fields 
\begin{equation}
\left[\phi_{\alpha} (x), \phi^{\dagger}_{\beta} (y)\right] = \delta_{\alpha\beta} \delta (x-y),\quad \alpha,\beta=1,2,\label{commutation}
\end{equation}
it can be checked that the nontrivial commutation relations for the creation and annihilation operators take the forms
\begin{equation}
\left[a (k), a^{\dagger} (k^{\prime})\right] = \delta (k - k^{\prime}) = \left[b (k), b^{\dagger} (k^{\prime})\right]
\end{equation}

Normally, in a relativistic theory, the time ordered propagator is the Feynman (causal) propagator. However, as a result of quantizing the theory in the wrong vacuum,  the propagator of the (relativistic) theory  becomes retarded (This is completely in spirit with the
Bethe ansatz methods in other systems) and has the form 
\begin{equation}
D(k)=\frac{i\left( k\!\!\!/+m\right) }{(k^{0}+i\varepsilon )^{2}-\left(
k^{1}\right) ^{2}-m^{2}}.  \label{propagator}
\end{equation}
As a result, the only diagrams that contribute to the two particle
scattering matrix are the bubble diagrams in Fig. 1. (This also clarifies
why the quartic part of the action is sufficient to study scattering of two
particles.) This simplification is automatic in non-relativistic models
where propagators are by definition retarded. However, in relativistic
models, this is achieved by quantizing the theory in the wrong vacuum (the
true vacuum being the one where all the negative energy states are filled).
\begin{figure}
[ptb]
\begin{center}
\includegraphics[
height=0.8311in,
width=2.8954in]
{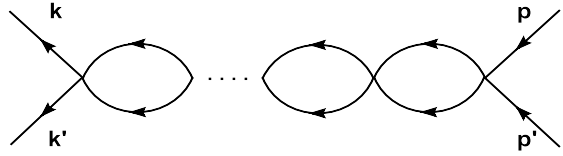}
\caption{The bubble diagrams contributing to the S-matrix}
\end{center}
\end{figure}

The calculation of the bubble diagrams is quite simple once the 1-loop
diagram is known. The interaction vertex in this theory can be written in a
tensor notation as 
\begin{equation}
igG=\frac{ig}{2}\left[ \gamma ^{0}\otimes \mathbf{1}+\mathbf{1}\otimes \gamma
^{0}-2\gamma ^{\mu }\otimes \gamma _{\mu }\right] ,  \label{Gtensor}
\end{equation}
and factoring out the vertex parts, the one loop contribution to the bubble
diagram has the form 
\begin{equation}
1\mathrm{-loop}=\int \frac{d^{2}q}{(2\pi )^{2}}\ D(p+p^{\prime }-q)\otimes
D(q),  \label{1loop}
\end{equation}
where $D(q)$ is the retarded propagator defined in \eqref{propagator}. For
positive energy external particles, this integral can be evaluated to yield
(for technical details, we refer the reader to \cite{Klose:2006dd}) 
\begin{equation}
1\mathrm{-loop}=\frac{1}{8m^{2}\sinh (\beta -\beta ^{\prime })}\left[ \left(
p\!\!\!/^{\prime }+m\right) \otimes \left( p\!\!\!/+m\right) +\left(
p\!\!\!/+m\right) \otimes \left( p\!\!\!/^{\prime }+m\right) \right] ,
\label{1loop2}
\end{equation}
where $\beta $ is the rapidity defined to be (see \eqref{rapidity}):
\begin{equation}
p^{0}=m\cosh \beta ;\text{ }p^{1}=m\sinh \beta .  \label{rapidity1}
\end{equation}
The general $n$-loop contribution to the S-matrix element , coming from the bubble diagrams, is then obtained
by raising the one loop result multiplied by the vertex functions raised to
the $n$th power and taking the matrix element between the incoming and the 
outgoing positive energy states $|p,p^{\prime }\rangle $ and $|k,k^{\prime
}\rangle $. The complete two particle S-matrix element corresponds to
summing over all loops and yields the result (because of energy-momentum conservation, $k=p,k^{\prime}=p^{\prime}$ or $k=p^{\prime},k^{\prime}=p$ as explained in \cite{Klose:2006dd}\textit{})
\begin{equation}
\langle k,k^{\prime }|\hat{S}|p,p^{\prime }\rangle =S(\beta ,\beta ^{\prime
})=\frac{1+i\lambda }{1-i\lambda },  \label{Smatrix_main}
\end{equation}
where 
\begin{equation}
\lambda \equiv g\ \frac{\cosh (\frac{\beta +\beta ^{\prime }}{2})-\cosh (\frac{
\beta -\beta ^{\prime }}{2})}{\sinh (\frac{\beta -\beta ^{\prime }}{2})}.
\label{lambda}
\end{equation}
Assuming the integrability of the model, and using the factorization
property of the S-matrix \cite{Zamolodchikov:1978xm}, one can now write a general $N$
-particle scattering S-matrix as a product of two-particle S-matrices. It is
worth noting here that the calculation of the S-matrix element involves an
inner product between the initial and the final positive energy states. As a
result, if the interactions of the incoming positive energy particles
generate an intermediate state that is orthogonal to the positive energy
states, the inner product with the positive energy outgoing states will not
see it. Therefore, the S-matrix calculation will not be sensitive to such an
issue. Normally, this is not an issue in other integrable models. However, as we will see in the next section, this question becomes quite
important in the diagonalization of the present Hamiltonian.

Once the two particle S-matrix is known, one can write the Bethe equations
as:
\begin{equation}
e^{i\sinh{\beta_{j}}L}=\underset{i\neq j}{\prod}S(\beta _{i},\beta _{j}).  \label{Bethe3}
\end{equation}
As we discussed in the introduction, the next logical step to carry out is to identify
the negative energy states and fill the Dirac sea with the purpose of
constructing the physical vacuum and the physical S-matrix. This well-known
procedure has been worked out in detail by Korepin for the fermionic
Thirring model \cite{Korepin:1979qq,Korepin:1979hg}. As the first step towards this, one needs to know
the explicit form of the two particle wave-function which will diagonalize the
Hamiltonian. This is the question that we take up in the next section.

\section{Diagonalization of the Hamiltonian}

Before proceeding, let us fix our conventions and notations. We have already mentioned that we use the Bjorken-Drell metric, $\eta^{\mu\nu} = (+,-)$ and although the explicit forms of the Dirac matrices are not relevant, for completeness we note that we use the following representations for the $\gamma$ matrices
\begin{equation}
\gamma ^{0}= \sigma_{1} =\left( 
\begin{array}{cc}
0 & 1 \\ 
1 & 0%
\end{array}%
\right),\text{ }\gamma^{1} = -i\sigma_{2}=\left( 
\begin{array}{cc}
0 & -1 \\ 
1 & 0%
\end{array}%
\right),\text{} \gamma_{5} = \gamma^{0}\gamma^{1} = \sigma_{3} =\left(
\begin{array}{cc}
1 & 0\\
0 & -1%
\end{array}\right).  \label{gammamat}
\end{equation}%
The two component spinor indices, when used, will be labelled by the beginning of the Greek alphabet $\alpha,\beta,\cdots = 1,2$. We use natural units where $\hbar = c = 1$.

The Hamiltonian density following from the quartic  action in (\ref{KZmain}) has the form
\footnote{As is conventionally done, we have omitted the term $m(\phi ^{\dagger}\phi ),$ which is proportional to the number operator, and commutes with the Hamiltonian.}
\begin{eqnarray}
\mathcal{H} &=& \mathcal{H}_{0} + \mathcal{H}_{I}\nonumber\\
& = & -i\phi ^{\dagger}\sigma _{3}\partial _{x}\phi +m\phi ^{\dagger}\sigma
_{1}\phi -g\left[ (\phi ^{\dagger}\phi )(\phi ^{\dagger}\sigma_{1}\phi )-(\phi ^{\dagger}\phi
)(\phi ^{\dagger}\phi )+(\phi ^{\dagger}\sigma _{3}\phi )(\phi ^{\dagger}\sigma _{3}\phi )
\right],  \label{KZHamiltonian}
\end{eqnarray}
which leads to the Hamiltonian
\begin{equation}
H =  H_{0} + H_{I} = \int dx\ \mathcal{H}.\label{hamiltonian}
\end{equation}
As we have noted earlier, this is the Hamiltonian that is responsible for the dynamics in the two particle sector. We note, that the structure of the
Hamiltonian (\ref{KZHamiltonian}) is reminiscent of the bosonic Thirring
model, namely, the last two terms in (\ref{KZHamiltonian}) describe exactly the interaction of the bosonic
Thirring model (while the first term in the interaction Hamiltonian violates Lorentz invariance). Thus, we can follow the standard procedure for the 
diagonalization of this Hamiltonian \cite{Thacker:1980ei,Korepin:1997bk}.

The theory, like all relativistic theories, has positive and negative energy solutions and let us construct the one particle states of the theory. As described in the last section, we will assume that the vacuum is annihilated by the field operator,
\begin{equation}
\phi (x) | 0\rangle = 0.
\end{equation}
Let us write the positive energy single particle state with momentum $k$ as
\begin{equation}
| \psi (k) \rangle _{(+)}  = \int dx\ \chi^{(+)}_{\alpha}(x|k)\phi _{\alpha}^{\dagger}(x) |0 \rangle .
\label{1partwf}
\end{equation}
Requiring this state to be an eigenfunction of the Hamiltonian \eqref{hamiltonian} with  positive energy,
\begin{equation}
H |\psi (k)\rangle_{(+)} = E_{k} |\psi (k)\rangle_{(+)},
\end{equation}
determines the spinor $\chi (x|k)$ to be (basically here one commutes the field variables to the right using \eqref{commutation} until it annihilates the vacuum \eqref{vacuum})
\begin{equation}
\chi^{(+)} (x|k) = \chi^{(+)} (x|\beta) = u_{+} (k) e^{ikx} = \sqrt{m}\left(\begin{array}{c}
e^{\frac{\beta}{2}}\\
e^{-\frac{\beta}{2}}
\end{array}\right) e^{i(m\sinh \beta)x},\label{positive}
\end{equation}
where we have used the rapidity variable defined earlier in \eqref{rapidity}, namely,
\begin{equation}
k = m \sinh \beta,\quad E_{k} = \sqrt{k^{2}+m^{2}}  = m \cosh \beta,
\end{equation}
to parameterize the solution. On the other hand, for the negative energy single particle state with momentum $k$, defining
\begin{equation}
|\psi (k)\rangle_{(-)} = \int dx\ \chi^{(-)}_{\alpha} (x|k)\phi_{\alpha}^{\dagger} (x) |0\rangle ,\label{1partwf1}
\end{equation}
and requiring that
\begin{equation}
H |\psi (k)\rangle_{(-)} = - E_{k} |\psi (k)\rangle_{(-)},
\end{equation}
we determine
\begin{equation}
\chi^{(-)} (x|k) = \chi^{(-)} (x|\beta) = u_{-} (k) e^{ikx} = \sqrt{m} \left(\begin{array}{c}
-e^{-\frac{\beta}{2}}\\
e^{\frac{\beta}{2}}
\end{array}\right) e^{i(m\sinh \beta) x}.\label{negative}
\end{equation}

The spinors in \eqref{positive} and \eqref{negative} are normalized to be consistent with the invariant normalization in \eqref{normalization} and, in fact, we can relate them to the covariant spinors in \eqref{covariant} and \eqref{covariant1} simply as $u_{+} (k) = u(k), u_{-} (k) = v (-k)$. Thus, these are, in fact, free single particle states, although they are eigenstates of the full Hamiltonian. This can be understood from the fact that the interaction Hamiltonian leads to a vanishing contribution to the single particle eigenvalue equation. We note now that if we identify the annihilation operators associated with the original field variables as
\begin{equation}
a_{\alpha} (k)=\int \frac{dx}{\sqrt{2\pi }}\ e^{-ikx}\phi _{\alpha}(x),\quad \alpha=1,2,  \label{aik}
\end{equation} 
then using \eqref{commutation}, we can obtain
\begin{equation}
\left[a_{\alpha} (k), \phi_{\beta}^{\dagger} (x)\right] = \delta_{\alpha\beta}\ \frac{e^{-ikx}}{\sqrt{2\pi}},\quad \alpha,\beta=1,2.\label{commutation1}
\end{equation}
Furthermore, let us define a new set of annihilation operators through a change of basis as
\begin{eqnarray}
A_{1} (k) &=& \frac{1}{\sqrt{2E_{k}}}\ u_{+,\alpha}^{\dagger} (k) a_{\alpha} (k) = \cos \theta _{k}a_{1}(k)+\sin \theta _{k}a_{2}(k) = \int \frac{dx}{\sqrt{4\pi E_{k}}}\ e^{-ikx} u_{+}^{\dagger} (k)\phi (x),  \notag \\
&&  \label{CAPaik} \\
A_{2} (k) &=& \frac{1}{\sqrt{2E_{k}}}\ u_{-,\alpha}^{\dagger} (k) a_{\alpha} (k) = -\sin \theta _{k}a_{1}(k)+\cos \theta _{k}a_{2}(k) =  \int \frac{dx}{\sqrt{4\pi E_{k}}}\ e^{-ikx} u_{-}^{\dagger} (k) \phi (x),  \notag
\end{eqnarray}
with $\theta_{k}$ satisfying the relation $k \tan 2\theta_{k} = m$. In fact, from the definition of the positive and the negative energy spinors in \eqref{positive} and \eqref{negative}, we note that
\begin{equation}
\cos\theta_{k} = \frac{e^{\frac{\beta}{2}}}{\sqrt{2\cosh \beta}},\quad \sin\theta_{k} = \frac{e^{-\frac{\beta}{2}}}{\sqrt{2\cosh \beta}}.\label{theta}
\end{equation}
In this new basis, it is easy to check that the free Hamiltonian is diagonal with the form
\begin{equation}
H_{0}=\int dk\ E_{k}\left( A_{1}^{\dagger} (k)A_{1} (k)-A_{2}^{\dagger} (k) A_{2} (k)\right),
\label{Hzero2}
\end{equation}
and we can identify the single particle states in \eqref{1partwf} and \eqref{1partwf1} with 
\begin{equation}
A_{1}^{\dagger} (k) |0\rangle = \frac{1}{\sqrt{2E_{k}}} |\psi (k)\rangle_{(+)} ,\quad A_{2}^{\dagger} (k)|0\rangle = \frac{1}{\sqrt{2E_{k}}} |\psi (k)\rangle_{(-)} .\label{physical1part}
\end{equation}
Namely, $A_{1}^{\dagger} (k)$ and $A_{2}^{\dagger} (k)$ can be thought of as the creation operators for single particle states with positive and negative energy respectively. It is worth noting here that the relations \eqref{CAPaik} can also be inverted to give
\begin{eqnarray}
a_{1} (k) & = & \cos\theta_{k} A_{1} (k) - \sin\theta_{k} A_{2} (k),\nonumber\\
a_{2} (k) & = & \sin\theta_{k} A_{1} (k) + \cos\theta_{k} A_{2} (k).\label{inversion}
\end{eqnarray}
More importantly, using the structures of the positive and the negative energy spinors, we can also invert \eqref{CAPaik} to determine the field expansion in terms of the new operators as
\begin{equation}
\phi (x) = \int \frac{dk}{\sqrt{4\pi E_{k}}}\ e^{ikx} \left(u_{+} (k) A_{1} (k) + u_{-} (k) A_{2} (k)\right).\label{expansion}
\end{equation}

To proceed with the diagonalization of the Hamiltonian in the two particle sector, we follow the standard procedure. However, to see clearly the effect of the interaction term in \eqref{KZHamiltonian} which violates Lorentz invariance, we put an arbitrary parameter in front of it and write
\begin{equation}
H_{I} = -g \int dx\left(\alpha \phi^{\dagger}\phi \phi^{\dagger}\sigma_{1}\phi-\phi^{\dagger}\phi \phi^{\dagger}\phi + \phi^{\dagger}\sigma_{3}\phi \phi^{\dagger}\sigma_{3}\phi\right),\label{int}
\end{equation}
where we note that for $\alpha =1$ we have the quartic Hamiltonian of the FR model, while for $\alpha = 0$ we get the bosonic Thirring model. Using the relations in \eqref{CAPaik} as well as the definition of the single particle states in \eqref{physical1part}, we note that in the absence of interactions, the two particle positive energy state with momenta $k_{1},k_{2}$ can be written as
\begin{eqnarray}
|k_{1},k_{2}\rangle _{(+)}
&=&A_{1}^{\dagger} (k_{1})A_{1}^{\dagger} (k_{2}) |0\rangle  \notag \\
&&  \label{wave1} \\
&=&\int \frac{dx_{1}dx_{2}}{4\pi \sqrt{E_{k_{1}}E_{k_{2}}}}\ e^{i(k_{1}x_{1}+k_{2}x_{2})} (\phi^{\dagger} (x_{1})u_{+} (k_{1})) (\phi^{\dagger} (x_{2})u_{+} (k_{2})) |0\rangle .
\notag
\end{eqnarray}%
In the presence of the quartic interactions, however, these would no longer correspond to the two particle (positive energy) eigenstates of the complete Hamiltonian. Thus, we generalize the definition of the two particle state with the ansatz
\begin{equation}
|\psi (k_{1},k_{2})\rangle_{(+)} = \int \frac{dx_{1}dx_{2}}{4\pi \sqrt{E_{k_{1}}E_{k_{2}}}}\  \chi (x_{1},x_{2}|k_{1},k_{2}) e^{i(k_{1}x_{1}+k_{2}x_{2})} (\phi^{\dagger} (x_{1})u_{+} (k_{1})) (\phi^{\dagger} (x_{2})u_{+} (k_{2})) |0\rangle ,\label{ansatz}
\end{equation}
where we assume
\begin{equation}
\chi (x_{1},x_{2}|k_{1},k_{2}) = 1 + i\lambda \varepsilon (x_{1}-x_{2}).\label{chi}
\end{equation}
Here $\lambda$ is a function of the momenta as well as the interaction strength (such that it vanishes in the absence of interactions) to be determined. One can, in principle, take a more general modification of the wavefunction, which we have done. However, the difficulty in diagonalizing persists nevertheless and, therefore, we discuss the issue with the conventional form of the generalization for the state. 

The calculation, which is slightly tedious, can be carried out in two steps. First, using \eqref{commutation} and moving the field operators to the right until they annihilate the vacuum \eqref{vacuum}, we obtain
\begin{equation}
H_{0} |\psi (k_{1},k_{2})\rangle_{(+)} = (E_{k_{1}}+E_{k_{2}}) |\psi (k_{1},k_{2})\rangle_{(+)} + |R\rangle ,\label{h0}
\end{equation}
where
\begin{equation}
|R\rangle  = -\frac{4\lambda \sin (\theta_{k_{1}} - \theta_{k_{2}})}{4\pi\sqrt{E_{k_{1}}E_{k_{2}}}} \int dx\ e^{i(k_{1}+k_{2})x} \phi_{1}^{\dagger} (x)\phi_{2}^{\dagger} (x) |0\rangle .\label{R}
\end{equation}
In carrying out the calculation, we have used the standard regularization $\delta (x) \varepsilon (x) = 0$. The action of the interaction Hamiltonian in \eqref{int} can similarly be calculated and leads to
\begin{eqnarray}
H_{I} |\psi (k_{1}, k_{2})\rangle_{(+)} &= & - \frac{g\alpha \sin (\theta_{k_{1}} + \theta_{k_{2}})}{4\pi\sqrt{E_{k_{1}}E_{k_{2}}}} \int dx\ e^{i(k_{1}+k_{2})x} \left(\phi_{1}^{\dagger} (x) \phi_{1}^{\dagger} (x) + \phi_{2}^{\dagger} (x) \phi_{2}^{\dagger} (x)\right)|0\rangle\label{extra}\\
&-&\!\!\frac{2g \left(\alpha \cos (\theta_{k_{1}} - \theta_{k_{2}}) - 2 \sin (\theta_{k_{1}}+\theta_{k_{2}})\right)}{4\pi\sqrt{E_{k_{1}}E_{k_{2}}}}\!\! \int dx\ e^{i(k_{1}+k_{2})x} \phi_{1}^{\dagger} (x) \phi_{2}^{\dagger} (x) |0\rangle .\nonumber
\end{eqnarray}
The two particle state \eqref{ansatz} would clearly be an eigenstate of the full Hamiltonian only if the sum of the terms in \eqref{R} and  \eqref{extra} cancel, which would also determine the parameter $\lambda$ as a function of the momenta and the interaction strength. However, in the present case, we see that since the structures of the terms in the two expressions are quite different, such a cancellation is not possible unless $\alpha = 0$ which would correspond to the bosonic Thirring model. In this case, we can determine  
\begin{equation}
\lambda = g\  \frac{\sin (\theta_{k_{1}}+\theta_{k_{2}})}{\sin (\theta_{k_{1}}-\theta_{k_{2}})} = - g \coth \frac{(\beta_{1}-\beta_{2})}{2},\label{bosonic}
\end{equation}
where, using \eqref{theta},  we have expressed the result in the rapidity variables in the last form. The S-matrix can now be written as
\begin{equation}
S = \frac{1 + i\lambda}{1-i\lambda},
\end{equation}
which is well known for the bosonic Thirring model. This result also holds for the fermionic Thirring model \cite{Bergknoff:1978bm,Bergknoff:1978wr}. However, we also note that for any nontrivial value of the Lorentz violating parameter $\alpha$, the two expressions in \eqref{R} and \eqref{extra} cannot be cancelled and hence the Hamiltonian cannot be diagonalized in the two particle sector.

To make connection with the field theoretic calculations in \cite{Klose:2006dd}, let us observe the following. We can rewrite the terms on the right hand side of  \eqref{extra} as
\begin{equation}
\int dx\ \frac{e^{i(k_{1}+k_{2})x}}{4\pi\sqrt{E_{k_{1}}E_{k_{2}}}} \left(|{\rm extra}\rangle -4g(\alpha\cos(\theta_{k_{1}}-\theta_{k_{2}}) -\sin(\theta_{k_{1}}+\theta_{k_{2}})) \phi_{1}^{\dagger} (x) \phi_{2}^{\dagger} (x) |0\rangle\right),
\end{equation}
where
\begin{equation}
|{\rm extra}\rangle = -g\alpha \left(\sin(\theta_{k_{1}}+\theta_{k_{2}}) (\phi_{1}^{\dagger}(x)\phi_{1}^{\dagger}(x)+\phi_{2}^{\dagger}(x)\phi_{2}^{\dagger}(x) - 2\cos(\theta_{k_{1}}-\theta_{k_{2}})\phi_{1}^{\dagger}(x)\phi_{2}^{\dagger}(x)\right)|0\rangle .
\end{equation}
Using the definitions in \eqref{wave1}, \eqref{CAPaik} and \eqref{commutation1}, it is now straight forward to check that
\begin{equation}
{}_{(+)}\langle k_{1},k_{2}|{\rm extra}\rangle = 0.
\end{equation}
Therefore, if we take the inner product of the sum of \eqref{R} and \eqref{extra} with a positive energy two particle state, the sum  will vanish provided (we remind the reader again that because of energy-momentum conservation, the momenta of the out states will coincide with a permutation of the momenta of the in states, as is well known \cite{Klose:2006dd})
\begin{equation}
\lambda = -g\ \frac{\alpha\cos(\theta_{k_{1}}-\theta_{k_{2}}) - \sin (\theta_{k_{1}}+\theta_{k_{2}})}{\sin (\theta_{k_{1}}-\theta_{k_{2}})} = g\ \frac{\alpha\cosh \frac{\beta_{1}+\beta_{2}}{2} - \cosh \frac{\beta_{1}-\beta_{2}}{2}}{\sinh \frac{\beta_{1}-\beta_{2}}{2}},\label{alpha}
\end{equation}
which, for $\alpha=1$ (FR model) reduces to the field theoretic result in \eqref{lambda}. This is indeed what we have tried to point out in the introduction. Namely, the S-matrix calculation involves calculating matrix elements and, consequently, is not sensitive to states orthogonal to the external states generated in the intermediate steps. These states, on the other hand, may be very important in the study of the diagonalization of the Hamiltonian. We will investigate this question in more detail in the next section.

\section{Solving the Boundary Condition} 

Although the analysis of the previous section makes it clear that the Hamiltonian is not diagonalizable in the presence of Lorentz violating terms, neither the origin of the problem nor the possible remedy is very clear. For that reason, let us analyze the reason for the difficulty in diagonalization from a different point of view. In this section, we will investigate the quantum mechanical problem associated with the diagonalization of the Hamiltonian in the two particle sector.

Let us write the quartic Hamiltonian for the system \eqref{KZHamiltonian}  (with the interaction in \eqref{int}) in the form
\begin{equation}
H=\int dx\left(-i\phi^{\dagger}\sigma _{3}\partial _{x}\phi +m\phi^{\dagger}\sigma
^{1}\phi -gV_{\alpha \beta ,\gamma \delta }\ \phi^{\dagger}_{\alpha
} \phi^{\dagger}_{\beta } \phi_{\gamma } \phi_{\delta }\right),  \label{H3}
\end{equation}
where  $V$ is the tensor related to $G$ defined in (\ref{Gtensor}) and has the 
explicit form (Unfortunately, the Lorentz violating parameter is called $\alpha$ just like the spinor index simply because we are running out of letters. However, we believe that there will be no confusion because of this.):
\begin{equation}
V = V_{\alpha \beta ,\gamma \delta }= (\sigma_{1}\otimes \sigma_{1}) G = \left( 
\begin{array}{crrc}
0 & \alpha & \alpha & 0 \\ 
\alpha & -2 & -2 & \alpha \\ 
\alpha & -2 & -2 & \alpha \\ 
0 & \alpha & \alpha & 0
\end{array}
\right).  \label{Gtensor2}
\end{equation}
We will use a tensor product notation in this section which makes the difficulty associated with the problem of diagonalization more transparent. Thus, the outer product of two positive energy spinors \eqref{positive} will be represented in this notation as a $4$- component column vector
\begin{equation}
U_{++,\alpha\beta} (k_{1},k_{2}) = u_{+,\alpha} (k_{1})u_{+,\beta} (k_{2}) = m\left(\begin{array}{c}
e^{\frac{\beta_{1}+\beta_{2}}{2}}\\
e^{\frac{\beta_{1}-\beta_{2}}{2}}\\
e^{-\frac{\beta_{1}-\beta_{2}}{2}}\\
e^{-\frac{\beta_{1}+\beta_{2}}{2}}
\end{array}\right),\label{tensor}
\end{equation}
(and so on for other spinors) and the contraction of the spinor indices would simply correspond to matrix products in this notation, which simplifies the calculations enormously.

Let us take a general ansatz for the two particle positive energy state as 
\begin{equation}
|\psi (k_{1},k_{2})\rangle_{(+)} = \int dx_{1}dx_{2}\ \chi_{\alpha\beta} (x_{1},x_{2}|k_{1},k_{2}) \phi^{\dagger}_{\alpha}(x_{1})\phi^{\dagger}_{\beta} (x_{2}) |0\rangle ,
\label{2part}
\end{equation}
where, unlike in \eqref{ansatz}, we have left the form of the wavefunction $\chi_{\alpha\beta} (x_{1},x_{2}|k_{1},k_{2})$ arbitrary, to be determined from the equations. Requiring that this state represents the two particle state of the complete Hamiltonian with positive energy of the form
\begin{equation}
H |\psi (k_{1},k_{2})\rangle_{(+)} = (E_{k_{1}}+E_{k_{2}}) |\psi (k_{1},k_{2})\rangle_{(+)},
\end{equation}
leads to (once again, one simply commutes the field variables to the right using \eqref{commutation} until they annihilate the vacuum \eqref{vacuum})
\begin{eqnarray}
& &\int dx_{1}dx_{2}\Big[\left( -i\left((\sigma _{3}\otimes \mathbf{1})\partial _{x_{1}} + (\mathbf{1}\otimes \sigma_{3})\partial_{x_{2}}\right) + m\left(\sigma_{1}\otimes \mathbf{1} + \mathbf{1}\otimes \sigma_{1}\right)
\right.  \notag \\
&&  \notag \\
&& 
\left. -(E_{k_{1}}+E_{k_{2}}) (\mathbf{1}\otimes\mathbf{1})  -2gV \delta (x_{1}-x_{2})\right) _{\alpha\beta,
\gamma\delta} \chi_{\gamma\delta} (x_{1},x_{2}|k_{1},k_{2})\Big]\phi_{\alpha}^{\dagger}(x_{1})\phi_{\beta}^{\dagger}(x_{2}) |0\rangle\nonumber\\
 & &\qquad =  0.\label{QMHam} 
\end{eqnarray}
Thus, we can interpret the expression in the large parenthesis (without the energy eigenvalue terms) as the quantum
mechanical Hamiltonian $H_{QM}$ in the two particle sector. To determine the two particle state \eqref{2part}, we need to solve the quantum mechanical equation
\begin{eqnarray}
& & \left[ -i\left((\sigma _{3}\otimes \mathbf{1})\partial _{x_{1}} + (\mathbf{1}\otimes \sigma_{3})\partial_{x_{2}}\right) + m\left(\sigma_{1}\otimes \mathbf{1} + \mathbf{1} \otimes \sigma_{1}\right)\right.
  \notag \\
&&  \notag \\
&&\quad \left. -(E_{k_{1}}+E_{k_{2}}) (\mathbf{1}\otimes\mathbf{1})  -2gV \delta (x_{1}-x_{2})\right] _{\alpha\beta,
\gamma\delta} \chi_{\gamma\delta} (x_{1},x_{2}|k_{1},k_{2}) = 0.\label{wfn}
\end{eqnarray}

Equation \eqref{wfn} is a first order equation with a delta potential whose coefficient (strength) has a rather nontrivial tensor structure. In this case, we expect that the wave function itself will be discontinuous across the boundary $x_{1}=x_{2}$. Away from the boundary (namely, in the region $x_{1}< x_{2}$ or $x_{1}> x_{2}$), the wave functions will correspond to the free two particle positive energy solutions whose coefficients must be determined by matching the discontinuity across the boundary.
Thus, let us choose the conventional general wave function of the form\footnote{For simplicity, we omit here the multiplicative factor of $\frac{1}{4\pi\sqrt{E_{k_{1}}E_{k_
{2}}}}$, which is not relevant for our analysis.}\cite{Korepin:1997bk}:
\begin{eqnarray}
\chi_{\alpha \beta}(x_{1},x_{2}|k_{1},k_{2})&=& e^{i(k_{1}x_{1}+k_{2}x_{2})} \left(1+i\lambda \varepsilon (x_{1}-x_{2})\right)U_{++,\alpha\beta}(k_{1},k_{2}) \nonumber\\
& & + e^{i(k_{1}x_{2}+k_{2}x_{1})}  \left(1-i\lambda \varepsilon
(x_{1}-x_{2})\right)U_{++,\alpha\beta}(k_{2},k_{1}),  \label{spinor2}
\end{eqnarray}
where $U_{++,\alpha\beta}(k_{1},k_{2})$ is the outer product of two positive energy solutions defined in \eqref{tensor}. Here $\lambda$ is assumed to be space independent and should, in principle, be determined from matching the boundary condition.  We comment here that the wave function in \eqref{spinor2} can be made even more general by giving $\lambda$ a nontrivial tensor structure. However, as we would see shortly, this does not help in the matching of the discontinuity across the boundary. 

It follows from \eqref{wfn} that at $x_{1}=x_{2}$, the discontinuity has to satisfy
\begin{eqnarray}
& & \left(\sigma_{3}\otimes \mathbf{1} - \mathbf{1}\otimes\sigma_{3}\right)_{\alpha\beta,\gamma\delta} (2\lambda) \left(U_{++,\gamma\delta}(k_{1},k_{2}) - U_{++,\gamma\delta}(k_{2},k_{1})\right)\nonumber\\
 & &\quad = 2g V_{\alpha\beta,\gamma\delta} \left(U_{++,\gamma\delta}(k_{1},k_{2}) + U_{++,\gamma\delta}(k_{2},k_{1})\right).\label{discontinuity}
 \end{eqnarray}
 As a result, the discontinuity relation takes the explicit form (in the outer product notation introduced in \eqref{tensor}),
 \begin{equation}
 4\lambda\left(\begin{array}{cc}
 0\\
 \sinh \frac{\beta_{1}-\beta_{2}}{2}\\
 \sinh \frac{\beta_{1}-\beta_{2}}{2}\\
 0
 \end{array}\right) = 2g \left(\begin{array}{cccc}
 \alpha \cosh \frac{\beta_{1}-\beta_{2}}{2}\\
 \alpha\cosh \frac{\beta_{1}+\beta_{2}}{2} - 2\cosh \frac{\beta_{1}-\beta_{2}}{2}\\
 \alpha\cosh \frac{\beta_{1}+\beta_{2}}{2} - 2\cosh \frac{\beta_{1}-\beta_{2}}{2}\\
 \alpha\cosh \frac{\beta_{1}-\beta_{2}}{2}
 \end{array}\right).\label{discontinuity1}
 \end{equation}
 Since the left and the right hand sides of the equation have quite different matrix structures, it is clear that the discontinuity relation cannot be satisfied and, therefore, a solution to \eqref{wfn} cannot be obtained for $\alpha\neq 0$. For $\alpha = 0$, for which the model corresponds to the bosonic Thirring model, the discontinuity condition can be satisfied and determines 
 \begin{equation}
 \lambda = -g \coth \frac{\beta_{1}-\beta_{2}}{2} ,
 \end{equation}
in agreement with the result \eqref{bosonic} in the operator method.
 
 At this point, one may wonder as to whether one cannot take a more general ansatz for the wave function in \eqref{spinor2}, for example, by giving a tensor structure to $\lambda$ allowing for a more compatible structure on the left hand side of \eqref{discontinuity1}. To investigate this question, let us note here that
 \begin{equation}
 \sigma_{3}\otimes\mathbf{1} - \mathbf{1}\otimes \sigma_{3} = \left(\begin{array}{rrrr}
 0 & 0 & 0 & 0\\
 0 & 2 & 0 & 0\\
 0 & 0 & -2 & 0\\
 0 & 0 & 0 & 0
 \end{array}
 \right),
 \end{equation}
 is a very special operator which, acting on a four component column vector, removes the top and the bottom elements. Therefore, even if we give a nontrivial tensor structure to $\lambda$ to allow for a more general component structure for the spinors, when the operator $(\sigma_{3}\otimes\mathbf{1} - \mathbf{1}\otimes \sigma_{3})$ acts on it, it would project out the top and the bottom elements and bring the spinor to the form on the left hand side of \eqref{discontinuity1}. Therefore, matching the discontinuity relation is the main problem because of which the Hamiltonian is not diagonalizable.

It is worth reflecting here on the connection between our analysis and the calculation of the S-matrix \cite{Klose:2006dd}. Let us note that although the discontinuity relation for the wave function \eqref{discontinuity} cannot be matched, if we take the inner product with the two particle positive energy state $U^{\dagger}_{++,\alpha\beta} (k_{1},k_{2})$, equation \eqref{discontinuity} can determine the parameter $\lambda$ to be
\begin{equation} 
\lambda =  g\  \frac{\alpha\cosh \frac{\beta_{1}+\beta_{2}}{2} -  \cosh \frac{\beta_{1}-\beta_{2}}{2}}{\sinh \frac{\beta_{1}-\beta_{2}}{2}},\label{KZalpha}
\end{equation}
which is what we have obtained in \eqref{alpha} and which, for $\alpha =1$, reduces exactly to the value calculated in \eqref{lambda} from field theoretic methods. To understand this better, let us note that the projection operator for the two particle positive energy states  with momenta $k_{1},k_{2}$ (as in \eqref{tensor}) can be easily computed to have the form
\begin{equation}
P_{++} (k_{1},k_{2}) = \frac{1}{4}\left( \begin{array}{cccc}
1 & e^{\beta_{2}} & e^{\beta_{1}} & e^{(\beta_{1}+\beta_{2})}\\
e^{-\beta_{2}} & 1 & e^{(\beta_{1}-\beta_{2})} & e^{\beta_{1}}\\
e^{-\beta_{1}} & e^{-(\beta_{1}-\beta_{2})} & 1 & e^{\beta_{2}}\\
e^{-(\beta_{1}+\beta_{2})} & e^{-\beta_{1}} & e^{-\beta_{2}} & 1
\end{array}\right),\label{projection}
\end{equation}
so that 
\begin{equation}
P_{++} (k_{1},k_{2}) U_{++} (k_{1},k_{2}) = U_{++} (k_{1},k_{2}).
\end{equation}
With this, we note that the column vector on the right hand side of \eqref{discontinuity1} can be uniquely decomposed into the sum
\begin{equation}
4g\left(\begin{array}{c}
0\\
\alpha \cosh \frac{\beta_{1}+\beta_{2}}{2} - \cosh \frac{\beta_{1}-\beta_{2}}{2}\\
\alpha \cosh \frac{\beta_{1}+\beta_{2}}{2} - \cosh \frac{\beta_{1}-\beta_{2}}{2}\\
0
\end{array}\right) + 2\alpha g\left(\begin{array}{c}
\cosh \frac{\beta_{1}-\beta_{2}}{2}\\
-\cosh \frac{\beta_{1}+\beta_{2}}{2}\\
-\cosh \frac{\beta_{1}+\beta_{2}}{2}\\
\cosh \frac{\beta_{1}-\beta_{2}}{2}
\end{array}\right),\label{decomposition}
\end{equation}
where the second vector is annihilated by the projection operator $P_{++}$ in \eqref{projection}. In the absence of this second term, the discontinuity relation \eqref{discontinuity1} (or \eqref{discontinuity}) can be solved and yield \eqref{KZalpha}. This is the reason why taking the inner product with positive energy states allows us to solve the discontinuity relation leading to the result from the field theoretic calculation. The important thing to note here is that in the perturbative calculation, one is evaluating matrix elements between positive energy states and, therefore, the calculation will not be sensitive to state vectors that are orthogonal to such states if they are generated in the intermediate steps.

There is a second way to look at this issue which is quite interesting. Let us note that  we can decompose the interaction potential \eqref{Gtensor2} uniquely as
\begin{equation}
V = \tilde{V} + V^{(0)},\label{decomposition1}
\end{equation}
where
\begin{eqnarray}
\tilde{V} &=& 2\left(\begin{array}{rrrr}
0 & 0 & 0 & 0\\
\alpha & -1 & -1 & \alpha\\
\alpha & -1 & -1 & \alpha\\
0 & 0 & 0 & 0
\end{array}\right),\nonumber\\
V^{(0)} &=& \alpha \left(\begin{array}{rrrr}
0 & 1 & 1 & 0\\
-1 & 0 & 0 & -1\\
-1 & 0 & 0 & -1\\
0 & 1 & 1 & 0
\end{array}\right).\label{decomposition2}
\end{eqnarray}
These matrices have the property that
\begin{eqnarray}
\tilde{V} U_{++} (k_{1},k_{2}) &=& 4m (\alpha\cosh \frac{\beta_{1}+\beta_{2}}{2} - \cosh \frac{\beta_{1}-\beta_{2}}{2}) \left(\begin{array}{c}
0\\
1\\
1\\
0
\end{array}\right),\nonumber\\
 V^{(0)} U_{++}(k_{1},k_{2}) &=& 2m\alpha\left(\begin{array}{rr}
\cosh \frac{\beta_{1}-\beta_{2}}{2}\\
-\cosh \frac{\beta_{1}+\beta_{2}}{2}\\
-\cosh \frac{\beta_{1}+\beta_{2}}{2}\\
\cosh \frac{\beta_{1}-\beta_{2}}{2}
\end{array}\right).
\end{eqnarray}
This is precisely the decomposition of the state vectors that we have discussed in \eqref{decomposition}. However, here the decomposition is in terms of the potential. What we see is that $V^{(0)}$ acting on a two particle positive energy state would lead to a state that is orthogonal to such a state. As a result, the S-matrix calculation, whether it is carried out with the vertex involving $\tilde{V}$ or the full vertex involving $V$, would lead to the same result since the extra terms generated by $V^{(0)}$ would drop out in the matrix element between positive energy states. On the other hand, from the point of view of diagonalizability, it is only $\tilde{V}$ that can be diagonalized and not the full $V$. We note that the extra term that is orthogonal to the positive energy states is proportional to the Lorentz violating parameter. Such a term is not present in the conventional (relativistic) integrable systems and this is a new feature in this model. This is the basis of our claim in the introduction that while a diagonalizable Hamiltonian leads to the S-matrix, having the S-matrix (say, from a field theoretic calculation) does not automatically imply that the Hamiltonian is diagonalizable.

\section{Diagonalizability and $\mathbf{PT}$ Symmetry}

The analysis of the last section is quite interesting and leads to the natural question as to what is the most general quartic Hamiltonian within this context that can be diagonalized and what is the corresponding S-matrix. Such an anlaysis would also determine the interaction Hamiltonian (potential) that would lead to the S-matrix calculated by Klose and Zarembo. This can be carried out systematically along the lines of discussion in the last section and we find that the most general Hamiltonian that can be diagonalized has the form
\begin{eqnarray}
H &=& H_{0} + H_{I} =  \int dx\left[-i \phi^{\dagger}\sigma_{3}\partial_{x}\phi + m \phi^{\dagger}\sigma_{1}\phi\right.\nonumber\\
 & & \quad \left. -g\left(\alpha \bar{\phi}\gamma^{0}\gamma^{\mu}\phi \bar{\phi}\gamma_{\mu}\phi + \beta \bar{\phi}\gamma^{\mu}\phi \bar{\phi}\gamma_{\mu} + \gamma (\bar{\phi}\phi \bar{\phi}\phi - \bar{\phi}\gamma_{5}\phi \bar{\phi}\gamma_{5}\phi)\right)\right].\label{generalH}
 \end{eqnarray}
 Here $\alpha,\beta$ and $\gamma$ are arbitrary real parameters (for the S-matrix to be unitary). We note that if $\alpha=\gamma=0$ and $\beta = -1$, this model reduces to the bosonic Thirring model that has been studied extensively. For $\alpha=\beta=0$ and $\gamma=-1$, this model corresponds to the bosonic chiral Gross-Neveu model which, to the best of our knowledge, has not been studied in the literature. Both these models involve Lorentz invariant interactions. Finally, the term with the parameter $\alpha$ clearly breaks Lorentz invariance, but does not coincide with the Lorentz violating term in \eqref{int}, rather generalizes it.
 
 The quantum mechanical potential, in this case, can be worked out in a straight forward manner (see last section) and has the form
 \begin{equation}
 \overline{V} = \left(\begin{array}{cccc}
 0 & 0 & 0 & 0\\
 \alpha & 2\beta & 2\gamma & \alpha\\
 \alpha & 2\beta & 2\gamma & \alpha\\
 0 & 0 & 0 & 0
 \end{array}\right).
 \end{equation}
 Substituting this potential into  \eqref{discontinuity1}, it is clear that the discontinuity relation can be satisfied with
 \begin{equation}
 \lambda = g\ \frac{\alpha\cosh \frac{\beta_{1}+\beta_{2}}{2} +(\beta+\gamma) \cosh \frac{\beta_{1}-\beta_{2}}{2}}{\sinh \frac{\beta_{1}-\beta_{2}}{2}}.\label{lambda1}
 \end{equation}
 It is important to note from \eqref{lambda1} that $\lambda$ is an antisymmetric function under the exchange of momenta, which is consistent for the Bose symmetry of the wave function.
 The S-matrix for this general diagonalizable model, then, follows from the standard relation
 \begin{equation}
 S = \frac{1+i\lambda}{1-i\lambda}.\label{S}
 \end{equation}
 It is now clear that the parameters $\alpha,\beta,\gamma$ must be real so that $\lambda$ is real and correspondingly the S-matrix is unitary.
 
 There are various special cases that one can study from \eqref{lambda1} and we list only three that we think are interesting.
 \begin{enumerate}
 \item First, if $\alpha = 0$ and $\beta+\gamma=-1$,
 \begin{equation}
 \lambda = -g\ \coth \frac{\beta_{1}-\beta_{2}}{2},
 \end{equation}
 which we recognize as the $\lambda$ for the bosonic Thirring model \eqref{bosonic}. However, what is striking here is that there is a one parameter family of interactions with $\alpha=0$ and $\beta+\gamma=-1$ which share the same value of $\lambda$ and, therefore, the S-matrix. In particular, we note that for $\beta=-1,\gamma=0$, the model corresponds to the bosonic Thirring model. On the other hand, for $\beta=0,\gamma=-1$, the model corresponds to the bosonic chiral Gross-Neveu model and we find that both the bosonic Thirring model as well as the bosonic chiral Gross-Neveu model (along with the one parameter family of interactions) share the same S-matrix element. To the best of our knowledge, this has not been recognized earlier. We also note that for $\alpha=0$, if $\beta+\gamma$ denotes an arbitrary constant (not equal to unity), this simply scales the coupling constant. However, for $\alpha=0$, if we also have $\beta=-\gamma\neq 0$, the S-matrix is trivial in spite of the fact that it is apparently an interacting theory. All of this can be understood as follows. In the outer product space of $2\times 2$ matrices, the completeness relation (Fierz identity) takes the form
 \begin{equation}
 \delta_{\alpha\beta}\delta_{\gamma\delta} = \frac{1}{2}\left[\delta_{\alpha\delta}\delta_{\gamma\beta} + (\sigma_{a})_{\alpha\delta}(\sigma_{a})_{\gamma\beta}\right],\quad a=1,2,3.\label{fierz}
 \end{equation}
 Using our convention for the gamma matrices in \eqref{gammamat}, we can also write this as
 \begin{equation}
 \delta_{\alpha\beta}\delta_{\gamma\delta} = \frac{1}{2}\left[\delta_{\alpha\delta}\delta_{\gamma\beta} + (\gamma^{\mu})_{\alpha\delta}(\gamma_{\mu})_{\gamma\beta} + (\gamma_{5})_{\alpha\delta}(\gamma_{5})_{\gamma\beta}\right],\quad \mu=0,1.\label{fierz1}
 \end{equation}
 Contracting \eqref{fierz1} with the bosonic fields $\bar{\phi}_{\alpha}\bar{\phi}_{\gamma}\phi_{\beta}\phi_{\delta}$ under the normal ordering sign, we obtain
 \begin{equation}
 \bar{\phi}\phi \bar{\phi}\phi - \bar{\phi}\gamma_{5}\phi \bar{\phi}\gamma_{5}\phi = \bar{\phi}\gamma^{\mu}\phi \bar{\phi}\gamma_{\mu}\phi ,\label{equivalence}
 \end{equation}
 which shows that in $1+1$ dimension, the bosonic Thirring interaction is equivalent to the bosonic chiral Gross-Neveu interaction, which explains the results noted above\footnote{For the massive fermionic Thirring model, it is well known that $:\bar{\psi}\gamma^{\mu}\psi \bar{\psi}\gamma_{\mu}\psi: \ \sim\ :\bar{\psi}\psi \bar{\psi}\psi:\ \sim\ \\:\bar{\psi}\gamma_{5}\psi \bar{\psi}\gamma_{5}\psi:$, which arises from the nilpotency of the fermionic fields.}. As a result of this equivalence, the last two terms in \eqref{generalH} can be combined into one. However, we keep them separate in the following discussion just for completeness.
 \item We note next that for $\alpha=1$ and $\beta+\gamma=-1$,
 \begin{equation}
 \lambda =  g\ \frac{\cosh \frac{\beta_{1}+\beta_{2}}{2} - \cosh \frac{\beta_{1}-\beta_{2}}{2}}{\sinh \frac{\beta_{1}-\beta_{2}}{2}}.
 \end{equation}
 This is exactly the $\lambda$ in \eqref{lambda} (and, therefore, the S-matrix) that has been calculated by Klose and Zarembo. We find that, although the FR Hamiltonian is not diagonalizable in the two particle sector, there exists a generalized interaction violating Lorentz invariance that can be diagonalized and leads to the same perturbative S-matrix as in \cite{Klose:2006dd}. In fact, as we have noted earlier, one can add any multiple of $V^{(0)}$ defined in \eqref{decomposition2} to this potential which would lead perturbatively to the same S-matrix element, but such Hamiltonians cannot be diagonalized. We note here that for $\alpha$ arbitrary with $\beta+\gamma=-1$, \eqref{lambda1} coincides with \eqref{alpha}.
 \item Finally, we note that if $\alpha=1$ and $\beta+\gamma=0$,
 \begin{equation}
 \lambda = g\ \frac{\cosh \frac{\beta_{1}+\beta_{2}}{2}}{\sinh \frac{\beta_{1}-\beta_{2}}{2}}.
 \end{equation}
 The integrability of this model, to the best of our knowledge, has not been studied earlier.
 \end{enumerate}
 
To summarize the results of the analysis of this section, we have determined the most general quartic Hamiltonian involving scalar fields that can be diagonalized. This involves three real parameters $\alpha,\beta,\gamma$ (actually two parameters if we use the equivalence in \eqref{equivalence}). The spectrum of $N$ particle states in such a system is real and is given by
\begin{equation}
E = m\sum_{i=1}^{N} \cosh \beta_{i},
\end{equation}
where the rapidities satisfy the Bethe equation:
\begin{equation}
e^{i\sinh{\beta_{j}}L}=\underset{i\neq j}{\prod}S(\beta _{i},\beta _{j}),  \label{Bethe4}
\end{equation}
and the S-matrix \eqref{S}, determined from \eqref{lambda1}, is unitary. However, if we look at the Lagrangian density or the Hamiltonian of the system, we find that it is not Hermitian. For example, we note that the Lagrangian density for the system has the form (we can, in principle, absorb the chiral Gross-Neveu interaction into the Thirring interaction using \eqref{equivalence}, but we keep them separate for completeness)
\begin{equation}
{\cal L} = \bar{\phi}(i\gamma^{\mu}\partial_{\mu} - m)\phi + g\left[\alpha\bar{\phi}\gamma^{0}\gamma^{\mu}\phi \bar{\phi}\gamma_{\mu}\phi +\beta \bar{\phi}\gamma^{\mu}\phi \bar{\phi}\gamma_{\mu}\phi + \gamma (\bar{\phi}\phi \bar{\phi}\phi - \bar{\phi}\gamma_{5}\phi \bar{\phi}\gamma_{5}\phi)\right],\label{lagrangiandensity}
\end{equation}
while the Hermitian conjugate is given by (up to a total derivative)
\begin{equation}
{\cal L}^{\dagger} = \bar{\phi}(i\gamma^{\mu}\partial_{\mu}-m)\phi + \left[\alpha\bar{\phi}\gamma^{0}(\gamma^{\mu})^{\dagger}\phi \bar{\phi}\gamma_{\mu}\phi + \beta \bar{\phi}\gamma^{\mu}\phi \bar{\phi}\gamma_{\mu}\phi + \gamma(\bar{\phi}\phi \bar{\phi}\phi - \bar{\phi}\gamma_{5}\phi \bar{\phi}\gamma_{5}\phi)\right],
\end{equation}
where we have used the reality of the parameters $\alpha,\beta,\gamma$. Thus, we see that the Lagrangian density is not Hermitian because of the interaction term violating Lorentz invariance. 

It is, therefore, surprising that the spectrum of the theory is real and the S-matrix is unitary. This can be understood from the fact that even though the theory (Lagrangian or the Hamiltonian) is not Hermitian, it is $PT$ symmetric. Such theories have been studied quite a lot in recent years from a variety of points of view \cite{Bender:2007nj,Bender:2005tb,Bender:1998gh,Dorey:2001hi,Bender:2002vv}. Here we describe its relevance within the context of this integrable model. First, we note that the simple two dimensional quantum mechanical example that is discussed extensively within the context of $PT$ symmetry is reminiscent of our positive and negative energy single particle solutions in \eqref{positive} and \eqref{negative}. Under parity transformation, $P$, we note that
\begin{equation}
x\rightarrow -x,\quad t\rightarrow t,\quad k\rightarrow -k,\quad E\rightarrow E.\label{parity}
\end{equation}
Parity is a linear operation and we can define its action on the field space by the relation
\begin{equation}
P \phi (x,t) P^{-1} = \eta_{P} \gamma^{0} \phi (-x,t),\quad P \bar{\phi} (x,t)P^{-1} = \eta_{P}^{*} \bar{\phi} (-x,t)\gamma^{0}.\label{parity1}
\end{equation}
Here $\eta_{P}$ is a phase denoting the intrinsic parity of the field. Time reversal, $T$, on the other hand is an antilinear operation defined on the coordinates by 
\begin{equation}
x\rightarrow x,\quad t\rightarrow -t,\quad k\rightarrow -k,\quad E\rightarrow E .\label{T}
\end{equation}
In the field space, the transformation can be described through the action
\begin{equation}
T \phi (x,t) T^{-1} = \eta_{T} C \gamma_{5} \phi (x, -t),\quad T \bar{\phi} (x, t)T^{-1} = \eta_{T}^{*} \bar{\phi} (x,-t) \gamma_{5} C^{-1},\label{T1}
\end{equation}
where $\eta_{T}$ is a phase and  $C$ denotes the (Dirac charge conjugation) matrix satisfying
\begin{equation}
C^{-1}\gamma^{\mu} C = - (\gamma^{\mu})^{T}.
\end{equation}
Although it is not necessary, we can choose the representation
\begin{equation}
C = -i\gamma^{1},\quad {\rm so\ that}\ \ C\gamma_{5}\gamma^{0} = \mathbf{1}.\label{id}
\end{equation}
With these transformations, it is easy to verify that under the parity transformation, the Lagrangian density is invariant, namely,
\begin{equation}
{\cal L} (x,t) \rightarrow {\cal L} (-x,t).
\end{equation}
On the other hand, since $T$ is an antilinear transformation, the Lagrangian density is $T$ invariant only for real parameters $\alpha,\beta,\gamma$ (which is the case we are considering), namely,
\begin{equation}
{\cal L} (x,t) \rightarrow {\cal L} (x,-t),
\end{equation}
only if $\alpha=\alpha^{*},\beta=\beta^{*},\gamma=\gamma^{*}$. Thus, for real parameters, we see that the theory is $PT$ symmetric although it is not Hermitian.

Let us next look at the behavior of the wave functions under this symmetry. We note that under the combined $PT$ transformation (see \eqref{parity}, \eqref{T} and remember that $T$ denotes an antilinear transformation)
\begin{equation}
e^{ikx}\rightarrow e^{ikx}.\label{phase}
\end{equation}
Similarly, if we choose the phase factors to be unity, namely $\eta_{P}=\eta_{T}=1$, under this combined operation 
\begin{equation}
u_{\pm} (k) \rightarrow C\gamma_{5}\gamma^{0} u_{\pm} (k) = \mathbf{1} u_{\pm} (k) = u_{\pm} (k),\label{spinortransformation}
\end{equation}
where we have used \eqref{id}. As a result, we see that the single particle positive and negative energy wave functions for the system in \eqref{positive} and \eqref{negative} are $PT$ symmetric.  Similarly, using the relations \eqref{phase} and \eqref{spinortransformation} in \eqref{wfn}, we find that the complete two particle wave function is also $PT$ symmetric (remember the anti-symmetry of $\lambda$ in the momenta). In other words, not only is the Hamiltonian of the theory $PT$ symmetric, but so are the wave functions of the theory. This implies that the theory is in an unbroken $PT$ symmetry phase which guarantees the reality of the spectrum as well as the unitarity of the S-matrix. This is indeed a novel demonstration of the relevance of $PT$ symmetry in an integrable system.

\section{Conclusion}

In this paper, we have studied in detail the Faddeev-Reshetikhin model, which is relevant in  the quantization of strings on $AdS_{5}\times S^{5}$, in the two particle sector. Although the S-matrix of the theory has been calculated using field theoretic methods, diagonalization of the Hamiltonian is  essential to carry out the Bethe ansatz analysis. We find that the quartic  Hamiltonian for this model is not diagonalizable in the two particle sector. We show this both in the operator description as well as in the description of the underlying quantum mechanical system. We trace the difficulty to the fact that the term in the interaction violating Lorentz invariance requires a discontinuity in the wave function that cannot be satisfied. On the other hand, if one takes the inner product of the discontinuity relation with positive energy states, the problematic term disappears leading to the correct S-matrix element calculated earlier. Further investigation shows that the interaction  (potential) generates intermediate states that are orthogonal to the positive energy out states and, therefore, cannot be observed in the S-matrix calculation (which involves calculating matrix elements), but are quite relevant in the diagonalization of the system. To the best of our knowledge, this is a new feature that has not been observed earlier in the study of integrable systems. It follows, therefore, that while the diagonalization of a Hamiltonian leads to the  S-matrix of the theory,  the knowledge of the S-matrix element (from, say, a field theoretic calculation) does not automatically guarantee the diagonalizability of the Hamiltonian of the system. We determine the most general Hamiltonian with quartic interactions that can be diagonalized as well as the associated S-matrix. Among various special cases, it also includes a generalized Hamiltonian that can be diagonalized with the S-matrix as calculated by Klose and Zarembo. We show that although this general Hamiltonian leads to a real spectrum and a unitary S-matrix, it is not Hermitian. However, we demonstrate that the theory is $PT$ symmetric and that wave functions are also invariant under $PT$. As a result, the theory is in an unbroken phase of $PT$ symmetry which is the reason for the reality of the spectrum as well as the unitarity of the S-matrix.

\section*{Acknowledgment}

One of us (A.D.) would like to thank Prof. J. Frenkel for many useful discussions and the Fulbright Foundation for financial support. A.M. would like to thank Aleksandr Pinzul for numerous insightful discussions. This work was supported in part by US DOE Grant number DE-FG 02-91ER40685, by CNPq and by FAPESP, Brazil. The work of A.M. was supported by the FAPESP grant No. 05/05147-3. The work of V.O.R. is supported by CNPq, FAPESP and PROSUL grant No. 490134/2006-8.

\bibliographystyle{JHEP3}
\bibliography{frmodel}

\end{document}